\documentstyle[12pt]{article}  
\begin{document}  \bibliographystyle{unsrt}

\noindent {\small From the preface of the Proceedings of the 4th
International Wigner Symposium, edited by N. M. Atakishiyev,
T. H. Seligman, and K. B. Wolf (World Scientific, Singapore, 1996)}.

\vspace{28mm}

\begin{center}
{\LARGE \bf Wigner's Sisters}\\[4mm]

Y. S. Kim\footnote{email: kim@umdhep.umd.edu}   \\
{\it Department of Physics, University of Maryland, \\College Park,
Maryland 20742, U.S.A.}
\end{center}

\vspace{2mm}

\begin{abstract}
Paul A. M. Dirac was a great physicist.  Wigner used to call
him {\em my famous brother-in-law}.  How did they become brothers-in-law?
Did these two great physicists have the same view toward physics?
\end{abstract}

I have been asked by the organizers of this Symposium to write about
Eugene Wigner's life.  Yes, he was a great physicist and was a great
human being.  I have been fortunate enough to have been associated
with him especially in his late years.  However, it will require years
of full-time research to write his biography if anyone decides to do so.
In the meantime, there is a an excellent book about him entitled {\em
Recollections of Eugene P. Wigner as told by Andrew
Szanton}~\cite{szanton92}.

At this time, I would like to define the scope of my knowledge about
Wigner by quoting a paragraph from what others are saying about me.  In
his review of the book entitled {\em Theory and Applications of the
Poincar\'e Group} which I wrote with Marilyn Noz~\cite{knp86}, Mariano
del Omo has the following paragraph~\cite{olmo88}.

\begin{quote}
E. P. Wigner's noteworthy paper [Ann. Math. {\bf 40}, 149-204 (1939)]
was the source of inspiration for the authors when writing this book.
There is also a remarkable trace of some of Dirac's papers in the book.
\end{quote}

According to this review, I am in a position to say something about
Wigner and his brother-in-law whose name was Paul Adrien Maurice Dirac.
When I was visiting Wigner frequently during the period 1985--90,
he had two sisters living in the United States.  They were all born in
Hungary in a well-to-do family.  His elder sister was in Binghamton
(New York), and his younger sister was and still is in Tallahassee
(Florida).  The elder sister's health was deteriorating, and Wigner
was always concerned about her and talking about her.  The younger
sister's name is Margit Dirac or Mrs. Paul A. M. Dirac.  She is known
as Manci in the physics community.  One day when I was in Wigner's
office at Princeton, he made a telephone call to Manci in order to say
``Happy Birthday'' to her.  After a brief talk in Hungarian, Wigner
laughed and told me Manci was complaining that his call disrupted her
shopping trip.  He then told me how she became Mrs. Dirac.

Manci was married to a very wealthy man in Hungary.  However, at that
time in Hungary or perhaps in other parts of the world, it was not
uncommon for a wealthy man to have several wives.  Manci's first husband
was a very handsome person in addition to being rich.  He was very
popular among women.  This was certainly not acceptable to her, and
she separated herself from him after having two children.  In 1934,
Manci visited her brother at Princeton.  He took her out for a dinner
at a restaurant called ``Annex''near the campus~\cite{annex}.  While
they were enjoying their dinner, Manci spotted a lonely-looking man
sitting at next table, and asked her brother who the man was.  Wigner
then looked at him, and he happened to be Paul A. M. Dirac.  They then
invited Dirac to join their table.  This is how Dirac became Wigner's
brother-in-law.

I met Mrs. Manci Dirac 1978 in Miami (Florida) while attending one of
the Coral Gables conferences.  I had a burning question to her husband,
and I abruptly joined the their ``husband-wife'' conversation in the
lobby of the hotel where the conference participants were staying.
After I finished the conversation with Paul Dirac, Manci asked me where
I came from originally.  I told her I came from Korea in 1954 right
after high-school graduation.  She then told me that I must have been
there during the Korean Conflict (1950--53).  After I said Yes to her,
she asked me how I felt about the result of the inconclusive war which
left the country divided.  It was quite clear to me that she was
extending to a man from Korea her sympathy toward Hungary which is also
prone to invasion and dominance by foreign powers, and I gave my
appropriate answer to her.  While she was talking, I also watched her
husband who was a great physicist.  He looked amused but did not show
any emotion.

I met Mrs. Dirac again in the fall off 1988 while I was visiting
Professor Wigner at Princeton.  He invited me to join a family dinner
consisting of his wife Eileen, his sister Manci, and himself.  We all
went to one of the ``Big Boy'' family restaurants in Trenton.  Assuming
that Manci knew about England because she lived there, I asked her a few
questions about the British prime ministers, particularly about Anthony
Eden who succeeded Winston Churchill but had to resign after the Suez
crisis in 1956.  Not many people talk about him these days.  She
explained to me the events during the Suez crisis like a history teacher,
and she had her own opinion about what happened and what did not happen
at that time.  I do not know her exact age, but she must have been about
eighty years old at that time.  She sounded like a professional lady of
my age.

Eugene Wigner also used to make his views known, and it is well known
that not everybody agreed with him on the issues having to do with the
communist world.  Yet, I have to point out that he told me many many
times Mikhail Gorbachev is a great man.  Wigner always wanted to live
peacefully with the people on the other side of the Iron Curtain.  He
had a distaste for the communist regime in Hungary, but his passion for
his native country was so strong that I had to contact the science
attache of the Hungarian Embassy in Washington.  As a result, the
Hungarian Ambassador invited Wigner to his residence during the spring
meeting of the American Physical Society (April 1988) held in Baltimore.
The Ambassador, presumably a member of the Hungarian Communist Party, was
kind enough to send his own limousine to Baltimore's convention center
where the APS meeting was held.  He later arranged Wigner's membership
in the Hungarian Academy of Sciences.  The Ambassador's name was Vencel
Hazi.  He did what he had to do, but I am still grateful to him.
Indeed, he was a very good communist.

Let us go back to Manci.  I told her that I was invited to the memorial
service held at Florida State University (Tallahassee) for Paul A. M.
Dirac in November of 1984 and I went there.  But I was not able to spot
her.  I asked her where she was at that time.  Her answer was that she
was so sad that she did not want to show her depressed face to others.
Indeed, she was talking like Queen Elizabeth or Margaret Thatcher.  It
is quite fortunate for the physics community that Manci took good
care of our respected Paul A. M. Dirac.  Dirac published eleven papers
during the period 1939--46.  It is not clear whether he knew Europe
went through World War II.  In either case, Dirac was able to maintain
his normal research productivity only because Manci was in charge of
everything else.

On the other hand, we were not fortunate enough to have Manci as a
physicist.  This is particularly so because there is a gap between
Dirac's approach and Wigner's approach to physics even though they had
the same ultimate goal in physics.  Manci could have filled this gap
if she had been born as a physicist.  What was then their common goal?
Dirac and Wigner both had a distaste for renormalization procedure, and
therefore they did not accept the present form of quantum field theory
as the ultimate theory.  Yet, both of them believed that the uncertainty
principle should someday be made consistent with special relativity if
not general relativity~\cite{noz88}.

During the period 1985--90, Wigner was keenly interested in approaching
this problem by constructing representations of the Poincar\'e group
using quantum phase-space distribution functions which are widely known
as Wigner functions.  Dirac, on the other hand, believed that fundamental
laws in physics should appear as beautiful mathematics.  His publication
list indicates clearly that he was quite fond of building relativistic
models using harmonic oscillators~\cite{dir43,dir45,dir63}.
I was indeed fortunate to be able to explains to Wigner what Dirac did,
and he used to enjoy listening to me.

Dirac wrote a number of papers on the Lorentz group.  His best known
paper on this subject is entitled {\em Forms of Relativistic Dynamics}
and is in the special issue of the {\em Reviews of Modern Physics}
dedicated to Einstein's 70th birthday in 1949~\cite{dir49}.  In this
paper, Dirac writes down the commutation relations, which he calls the
Poisson brackets, for the generators of the Poincar\'e group, and states
that {\em the problem of finding a new dynamical system reduces to the
problem of finding a new solutions of these equations.}  This is exactly
what Wigner proposed in his 1939 paper on the {\em Inhomogneous Lorentz
Group}~\cite{wig39}.   Dirac's {\em instant form} and {\em front form}
can be connected to Wigner's $O(3)$-like and $E(2)$-like little groups
for massive and massless particles respectively~\cite{knp86}.  As I said
earlier in this report, I had a ``burning question'' to Dirac in 1978
simply because I wanted to understand Dirac's 1949 paper~\cite{dir49}
in terms of Wigner's representation theory.  Dirac of course gave me his
clear answers in terms of what he said in his own papers, but he was not
familiar with the papers written on the same subject by his {\em famous
brother-in-law}.

It is somewhat frustrating to note that these two {\em great
brothers-in-law} did not have much communication with each other in
physics.  On the other hand, I was able to find many homework problems
from this gap, and this is why I was able to write my first book. This
will explain why Del Olmo made a remark about Dirac's influence on my
book with Noz~\cite{knp86,olmo88}.  But this story is not restricted
to me or to my book.  The communication gap between these two great
physicists offers a great challenge to many young physicists.  Try to
establish a bridge between Dirac and Wigner.  It may become a very
profitable enterprise.


\begin{thebibliography}{99}

\bibitem{szanton92}
A. Szanton, {\em The Recollection of Eugene P. Wigner} (Plenum, New York,
1992).

\bibitem{knp86}
Y. S. Kim and M. E. Noz, {\em Theory and Applications of the Poincar\'e
Group} (Reidel, Dordrecht, 1986).

\bibitem{olmo88}
M. A. del Omo, {\em Mathematical Review} {\bf 88a}, 160 (1988).

\bibitem{annex}
The Annex restaurant is located on Nassau Street across from the
Firestone Library of Princeton University.

\bibitem{noz88}
M. E. Noz and Y. S. Kim, {\em Special Relativity and Quantum Theory},
Edited Volume consisting of Wigner's papers, Dirac's papers, and others
(Kluwer, Dordrecht, 1988).

\bibitem{dir43}
P. A. M. Dirac, {\em Quantum Electrodynamics, Comm. Dublin Inst.
Adv. Stud. ser. A, No. 1} (1943).

\bibitem{dir45}
P. A. M. Dirac, {\em Proc. Roy. Soc. (London)} {\bf A183}, 284 (1945).

\bibitem{dir63}
P. A. M. Dirac, {\em J. Math. Phys.} {\bf 4}, 901 (1963).

\bibitem{dir49}
P. A. M. Dirac, {\em Rev. Mod. Phys.} {\bf 21}, 392 (1949).

\bibitem{wig39}
E. P. Wigner, {\em Ann. Math.} {\bf 40}, 149 (1939).

\end{thebibliography}
\end{document}